\documentclass[final]{ias2}

\usepackage{graphicx}
\usepackage{multirow}
\usepackage{array}

\usepackage{hyperref}

\newcommand{\La}{\langle}
\newcommand{\Ra}{\rangle}

\begin{document}

\markboth{Hydrodynamic Modeling for Relativistic Heavy Ion Collisions at RHIC and the LHC}{H.Song}

\title{Hydrodynamic Modeling for Relativistic Heavy Ion Collisions at RHIC and the LHC}

\author[sin]{Huichao Song}
\email{huichaosong@pku.edu.cn}
\address[sin]{Department of Physics and State Key Laboratory of Nuclear Physics and Technology, Peking
University, Beijing 100871, China}

\begin{abstract}
In this article, we briefly review recent progress on hydrodynamic modeling and its implementations to relativistic heavy-ion collisions at RHIC and the LHC. The related topics include: 1) initial state fluctuations, final state correlations and event-by-event hydrodynamics, 2) extracting the QGP shear viscosity from flow data, 3) flow and hydrodynamics in 5.02 A TeV p+Pb collisions.
\end{abstract}

\keywords{The quark gluon plasma, hydrodynamics, flow, viscosity}

\pacs{25.75.-q, 12.38.Mh, 25.75.Ld, 24.10.Nz}

\maketitle


\section{Introduction}
\quad The quark gluon plasma (QGP) has been created in high energy heavy ion collisions at Relativistic Heavy-Ion Collider (RHIC) and the Large Hadron Collider the (LHC)~\cite{Rev-Arsene:2004fa,Gyulassy:2004vg,Muller:2006ee}. The observations of strong collective flow and the successful descriptions from hydrodynamics have established that the QGP is strongly coupled and behaves like an almost perfect liquid~\cite{Gyulassy:2004vg,Muller:2006ee,reviews,Heinz:2013th,Gale:2013da}. The specific QGP shear viscosity $(\eta/s)_{QGP}$ is one of the key quantities to evaluate the strongly coupled nature of the QCD matter and to answer how perfect is the QGP fluid.

Early viscous hydrodynamic calculations revealed that the QGP shear viscosity suppresses the fluid anisotropy and can be extracted from the flow data measured in experiments~\cite{Heinz:2013th,Gale:2013da,Teaney:2009qa,Romatschke:2009im,Romatschke:2007mq,Song:2007fn,Dusling:2007gi,Molnar:2008xj,Bozek:2009dw,Chaudhuri:2009hj,Schenke:2010rr}. However, hadronic chemical compositions and off-equilibrium kinetics also significantly influence the development of flow~\cite{Kolb:2002ve,Huovinen:2007xh,Hirano:2005wx}, leading to large contaminations for the extracted value of the QGP shear viscosity. For a better description of the hadronic matter, hybrid models have been developed by several groups through coupling viscous hydrodynamics for the QGP fluid expansion with a hadron cascade model to describe the microscopic evolution and decoupling of the hadronic matter~\cite{Song:2010aq,Soltz:2012rk,Ryu:2012at,Karpenko:2013ama}.

Using VISHNU hydrid model, the specific QGP shear viscosity has been semi- quantitatively extracted from the elliptic flow data with smooth initial conditions generated from MC-Glauber and MC-KLN models~\cite{Song:2010mg,Song:2011hk}. The initial state fluctuations in these two models are constructed from the fluctuations of nucleon positions~\cite{Alver:2006wh,Drescher:2006ca,Hirano:2009ah}. Around 2012, other sources of quantum fluctuations have been further explored, including color charge fluctuations~\cite{Dumitru:2012yr,Muller:2011bb,Moreland:2012qw,Schenke:2012hg,Gale:2012rq}, initial flow fluctuations and longitudinal fluctuations~\cite{Pang:2012he}, etc. Meanwhile, new flow data related to higher order flow harmonics have been measured~\cite{Adare:2011tg,Adamczyk:2013waa,ALICE:2011ab,Aamodt:2011by,ATLAS:2012at,Jia:2012ve,Jia:2012sa}, which have been extensively investigated by event-by-event hydrodynamic simulations~\cite{Schenke:2012hg,Gale:2012rq,Pang:2012he,Petersen:2010cw,Qin:2010pf,Holopainen:2010gz,Qiu:2011iv,Qiu:2012uy,Luzum:2013yya}. These experimental and theoretical progress provide new opportunities for a tight constrain of the initial conditions and an accurate extraction of the QGP shear viscosity in the near future~\cite{Song:2012ua}. In this article, we will review recent progress on hydrodynamic modeling, its investigation on the flow data and the extraction of the QGP shear viscosity in relativistic heavy ion collisions at RHIC and the LHC.

\section{Hydrodynamic modeling - a short introduction}

\subsection{Viscous hydrodynamics}

\quad Relativistic hydrodynamics is a macroscopic tool to simulation the QGP fireball evolution and to describe and predict the soft particle physics in relativistic heavy ion collisions~\cite{reviews}. It is based on the conservation laws of energy, momentum and net charge current. The equations are written as:
\begin{subequations}
\begin{eqnarray}
\partial_\mu T^{\mu \nu}(x)=0\, , \qquad \qquad \partial_\mu N^{\mu} (x)=0\, .  \nonumber \hspace{3cm}
\end{eqnarray}
\end{subequations}

Ideal hydrodynamics assumes local equilibrium, which expresses the energy momentum tensor and the net baryon charge current as: $T^{\mu \nu}=(e+p) u^{\mu}u^{\nu}-p g^{\mu\nu}$ and $N^{\mu}=nu^{\mu}$. Here, the 14 independent variables in $T^{\mu\nu}$ and $N^{\mu}$  reduce to  6 unknowns (1 each for the energy density $e$, pressure $p$ and net baryon density $n$, and 3 independent components in the fluid four velocity $u^\mu$). After inputting the equation of state (EoS) $p=p(n,e)$, the system is closed. The set of equations can be solved numerically with properly chosen initial conditions~\cite{reviews}.

Viscous hydrodynamics works for a near equilibrium system. In the Laudau frame, $T^{\mu\nu}$ and $N^{\mu}$ are expressed as: $T^{\mu \nu}=(e+p+\Pi) u^{\mu}u^{\nu}-(p +\Pi)g^{\mu\nu}+\pi^{\mu \nu}$, $N^{\mu}=nu^{\mu}+V^{\mu} $.  Here, $\pi^{\mu \nu}$ is the shear stress tensor, $\Pi$ is the bulk pressure and $V^{\mu}$ is the baryon flow. At top RHIC and the LHC energies, the net baryon density $n$ is negligible and $V^{\mu}$ is assumed to be zero. The additional evolution equations for the viscous terms can be obtained from the 2nd law of thermal dynamics or from kinetic theory, which have the following forms~\cite{Israel:1976tn,Muronga:2004sf}:
\begin{eqnarray}
\label{eq2}&&\Delta^{\mu\alpha}\Delta^{\nu\beta} D\pi_{\alpha\beta}
=-\frac{1}{\tau_{\pi}}(\pi^{\mu\nu}{-}2\eta\sigma^{\mu\nu})
 -\frac{1}{2}\pi^{\mu\nu} \frac{\eta T}{\tau_\pi}
       \partial_\gamma\left(\frac{\tau_\pi}{\eta T}u^\gamma\right),\nonumber
\\ \label{eq3}
        &&D \Pi
=-\frac{1}{\tau_{\Pi}}(\Pi+\zeta \theta)
 -\frac{1}{2}\Pi\frac{\zeta T}{\tau_\Pi}
       \partial_\gamma\left(\frac{\tau_\Pi}{\zeta T}u^\gamma\right).\nonumber
\end{eqnarray}
where $D=u^\mu \partial_\mu $, $\Delta^{\mu\nu} =g^{\mu\nu}{-}u^\mu u^\nu$, $\sigma^{\mu\nu}=\nabla ^{\La \mu}u
^{\nu\Ra}$, and $\La ... \Ra$ denote the symmetric and traceless projection, orthogonal to fluid four velocity $u^\mu$ . $\eta$ is the shear viscosity, $\zeta$ is the bulk viscosity, and $\tau_{\pi}$ and $\tau_{\Pi}$ are the corresponding relaxation times\footnote{Here we concentrate on the 2nd order viscous hydrodynamics for a near equilibrium system with isotropic momentum distributions. For very early fluid expansion, one needs to implement anisotropic viscous hydrodynamics using a reorganized formalism to incorporate the large momentum anisotropy. The recent developments can be found in~\cite{Martinez:2010sc,Florkowski:2010cf,Bazow:2013ifa}.}.

The EoS is one of the key inputs in hydrodynamic simulations. The state-of-Art EoS used by many group is s95p-PCE~\cite{Huovinen:2009yb,Shen:2010uy}. It matches the recent lattice EoS with a partially chemical equilibrium hadronic EoS at 165 MeV, which corresponds to the chemical freeze-out temperature measured at RHIC~\cite{BraunMunzinger:2001ip}.

At the starting time $\tau_0$, hydrodynamic simulations require initial entropy/energy density and initial flow velocity to start the evolutions. These initial profiles can be provided by different initial condition models or by pre-equilibrium dynamics. This will be further discussed in Sec.~3.1. The free parameters in initialization models are constrained by several observables measured in experiments~\cite{reviews}.

To obtain final hadrons, pure hydrodynamic simulations assume free hadron resonances directly emit from the fluid along a decoupling surface. The Cooper-Frye formula~\cite{Cooper-Frye} is then implemented to calculate the particle momentum distributions, which is followed by a resonance delay routine to generate final stable hadrons. The decoupling surface can be defined by a constant temperature or energy density or other kinetic variables\cite{reviews,Teaney:2009qa}. For the scenario of constant temperature decoupling, $T_{dec}$ is generally set to 100-130 MeV, depending on the EoS and other hydrodynamic inputs, to allow for sufficient evolution time to build up enough radial flow to fit the slopes of the $p_T$ spectra~\cite{reviews}.

To simplify numerical simulations, many viscous hydrodynamic calculations assume a specific velocity profile $v_z = z/t$ along the beam directions (Bjorken approximation). This leads to a longitudinal boost invariance and reduces the (3+1)-d hydrodynamics to (2+1)-d hydrodynamics. The (2+1)-d codes have been developed by several groups since 2007~\cite{Romatschke:2007mq,Song:2007fn,Dusling:2007gi,Molnar:2008xj,Bozek:2009dw,Chaudhuri:2009hj}. Many of these independent developed codes have passed the standard code verifications within the TECHQM collaboration~\cite{TECHQM}. Recently, several groups~\cite{Schenke:2010rr,Bozek:2011ua,Vredevoogd:2012ui,Nonaka:2013uaa,DelZanna:2013eua,Karpenko:2013wva} have further developed (3+1)-d viscous hydrodynamics without longitudinal boost invariance. A recent comparison between the 2+1-d and 3+1-d codes showed that the realistic longitudinal expansion only slightly affects the flow profiles at mid-rapidity~\cite{ChunNote}. One can still safely investigate the soft particle physics at mid-rapidity using a 2+1-d code.

\subsection{Hybrid models}
\quad A hybrid model matches hydrodynamic descriptions of the expanding QGP to microscopic Boltzmann simulations of the evolving hadronic matter. The transition between models is realized by a monte-carlo event generator, which transforms the hydrodynamic output into individual hadrons for succeeding hadron cascade propagations. Early hybrid models couple ideal hydrodynamics (in 1+1-d 2+1-d and 3+1-d versions) with a hadron cascade model~\cite{Bass:2000ib,Teaney:2000cw,Nonaka:2006yn,Hirano:2005wx}. A comparison between the hybrid model and pure ideal hydrodynamics with a partially chemical equilibrium EoS showed that the late hadronic evolution is highly viscous or even far from equilibrium, leading to an O(30\%) reduction of the elliptic flow~\cite{Hirano:2005wx}.  This motivated the development of viscous hydrodynamics + hadron cascade hybrid model for an accurate extraction of the QGP viscosity from the flow data. In 2010, the OSU-LBL group developed VISHNU hybrid model that couples 2+1-d viscous hydrodynamics with the UrQMD hadron cascade~\cite{Song:2010aq}. It was found that the longitudinal boost invariance are well kept at mid-rapidity after the 3+1-d hadron cascade evolution~\cite{Song:2010aq} (Please also refer to~\cite{Soltz:2012rk} for the 2+1-d hybrid code developed in Livermore). After 2012, full 3+1-d hybrid models were individually developed by the McGill and Frankfurt groups through connecting 3+1-d viscous hydrodynamics with UrQMD~\cite{Ryu:2012at,Karpenko:2013ama}.

Comparing with the pure hydrodynamics, the hybrid approach imprints dynamical chemical and thermal freeze-out for varies hadron species. The off-equilibrium hadronic evolution brings additional viscous suppressions for the anisotropy flow, which become more and more important at lower collision energies~\cite{Song:2010aq,Karpenko:2013ama}.  It improves the descriptions of $v_2$ mass-splitting between pions and protons through the microscopic hadronic rescatterings that rebalance the generation of radial and elliptic flow~\cite{Song:2013qma,Heinz:2011kt}. The baryon-antibaryon ($B-\bar{B}$) annihilations in the hadronic evolution play an important role for the hadron yields, which reduce the proton and antiproton multiplicity by O(30\%) and help to achieve better descriptions of the experimental data~\cite{Song:2013qma,Song:2012ua}. In contrast, the $B-\bar{B}$ annihilations scenario could not be easily adapted in pure hydrodynamic simulations or in the statistical model.

\subsection{Single shot simulations vs. event-by-event simulations:}

\quad The nucleons inside each colliding nuclei fluctuate event-by-event, leading to fluctuating initializations for the succeeding QGP fireball evolutions~\cite{Alver:2006wh,Miller:2003kd,Alver:2008zza}. One can directly input the fluctuating initial profiles for individually evolved hydrodynamic simulations and then average the results (\emph{event-by-event simulations}). One can also obtain a smooth initial condition through averaging over large numbers of events at the beginning and then use it for one hydrodynamic simulation (\emph{single-shot simulation}).  Before 2010, most of the hydrodynamic calculations concentrated on single-shot simulations for the ease of numerical implementations. Recently, initial state fluctuations and final state correlations became hot research topics. Event-by-event simulations have been further developed to investigate the hydrodynamic response of the initial fluctuations and to study the corresponding experimental data~\cite{Schenke:2012hg,Gale:2012rq,Pang:2012he,Petersen:2010cw,Qin:2010pf,Holopainen:2010gz,Qiu:2011iv,Qiu:2012uy}.

For computing efficiencies, one could calculate the elliptic and triangular flow ($v_2$ and $v_3$) using single-shot simulations with smooth initial conditions obtained through averaging thousands of fluctuating events with reaction plane or participant plane aligned. A comparison with the event-by-event simulations showed that the differences are less than 10\% for $v_2$ and $v_3$~\cite{Qiu:2011fi}. However, the higher order flow harmonics, $v_4, v_5$ and $v_6$, can not be reliably computed through the single shot simulations due to the coupling between modes. Furthermore, some of the flow measurements, such as event-by-event $v_n$ distribution~\cite{Jia:2012ve} and the event plan correlations between flow angles~\cite{Jia:2012sa}, can only be investigated within the framework of event-by-event simulations.

\section{Initial state fluctuations and final state correlations}

\subsection{Fluctuating initial conditions}

\quad MC-Glauber and MC-KLN models~\cite{Alver:2006wh,Drescher:2006ca} are two commonly used initializations for hydrodynamic simulations. Although the treated degrees of freedom in these two models are nucleons and gluons respectively, the initial state fluctuation are both constructed through the positions fluctuations of nucleons inside each colliding nuclei. Since 2012, the quantum fluctuations of color charges have been further investigated.  IP-Glama model~\cite{Schenke:2012hg} combines the IP-Sat model for high energy nuclei/nucleon wave-functions with the classical Yang-Mill's dynamics for the pre-equilibrium glasma evolution. It includes both nucleon position and color charge fluctuations, which gives moderately modified event averaged initial eccentricities $\varepsilon_n$, but obviously different event-by-event $\varepsilon_n$ distributions compared with the ones from MC-Glauber and MC-KLN~\cite{Schenke:2012hg,Gale:2012rq}. Within the framework of MC-KLN model, correlated initial fluctuations have been constructed by the OSU group~\cite{Moreland:2012qw} using a covariance function derived in~\cite{Muller:2011bb}. It was found that, with a realistic correlation length, the additional correlated gluon filed fluctuations increase the eccentricities $\varepsilon_n$ by less than 10\%. Ref.~\cite{Dumitru:2012yr} studied the local multiplicity fluctuations for initial gluon productions through implementing the negative binomial distributions to the $k_T$ factorization formulism of the Color Glass Condensate. This model gives an initial eccentricity $\varepsilon_2$ close to that from MC-KLN, but obvious larger $\varepsilon_3 -\varepsilon_5$ than the MC-KLN ones.

Dynamical models, such as URQMD~\cite{Bass:1998ca}, EPOS~\cite{EPOS}, AMPT~\cite{Lin:2004en} and IP-Glama~\cite{Schenke:2012hg}, can provide pre-equilibrium dynamics for the succeeding hydrodynamic evolutions. After matching the energy-momentum tensor between models at a switching time, both fluctuating initial energy density and fluctuating initial flow are obtained. While most of the investigations constrained the initial fluctuations to the transverse plane, Pang and his collaborators studied the longitudinal fluctuations through combining event-by-event 3+1-d ideal hydrodynamics with AMPT pre-equilibrium dynamics. They found that the evolving longitudinal hot spots dissipate part of the transverse energy, leading to a suppression of flow anisotropy in the transverse directions~\cite{Pang:2012he}.

\subsection{Collective flow and event-by-event hydrodynamic simulations}

\quad The fluctuating initial conditions results in the presence of odd flow harmonics~\cite{Alver:2010gr,Alver:2010dn} and a finite elliptic flow at zero impact parameter~\cite{Alver:2006wh} in relativistic heavy ion collisions. These quantities were once predicted to be zero by single-shot hydrodynamics with smooth and symmetric initial conditions. Recently, triangular flow and other higher order flow harmonics have been measured at RHIC~\cite{Adare:2011tg,Adamczyk:2013waa} and the LHC~\cite{ALICE:2011ab,Aamodt:2011by,ATLAS:2012at}, followed by intensive theoretical investigations~\cite{Schenke:2012hg,Gale:2012rq,Pang:2012he,Petersen:2010cw,Qin:2010pf,Holopainen:2010gz,Qiu:2011iv,Qiu:2012uy}.  Event-by-event hydrodynamic simulations revealed that the flow from $v_2$ to $v_5$ are all suppressed by the shear viscosity. Higher flow harmonics show more sensitivity to $\eta/s$ since higher order granularities are smeared out early during the QGP evolution~\cite{Schenke:2012hg,Gale:2012rq}. Therefore, one can extract the QGP shear viscosity from the flow harmonics at different orders using event-by-event hydrodynamic simulations.

These event-averaged flow harmonics $v_n$ mainly reflect the hydrodynamic response of the event-averaged initial eccentricity $\varepsilon_n$~\cite{Retinskaya:2013gca}. On an event-by-event basis, the ATLAS collaboration measured the $v_n$ distributions (n=2...4) in 2.76 A TeV Pb+Pb collisions~\cite{Jia:2012ve}. Event-by-event hydrodynamic simulations showed that, after rescale the distributions by the corresponding mean values, the $v_n$ distributions mostly follow the $\varepsilon_n$ distributions which is independent of the hydrodynamic evolution details~\cite{Gale:2012rq}. The measured $\varepsilon_n$ distributions thus can be used to roughly constrain the initialization models. Both MC-Glauber and MC-KLN models are disfavored by the measured $v_n$ distributions since none of them gives a rescaled $\varepsilon_n$ distributions that successfully reproduce the $v_n$ distributions at all centralities. The situations is even worse for n=2 from semi-central to peripheral collisions~\cite{Jia:2012ve}. Ref~\cite{Gale:2012rq, Gale:2013da} showed that the $\varepsilon_n$ distributions from IP-Glasma nicely overlap with the $v_n$ distributions for the selected centrality bins, except for the tails of the $\varepsilon_4$ distributions. The non-liner  hydrodynamic evolution couples different harmonic modes. After event-by-event hydrodynamic simulations,
an improved descriptions of the $v_4$ distributions, together with nice fit of the measured $v_2$ and $v_3$ distributions, is achieved.

More fluctuation information is provided by the event plan correlations among flow angles, which have been recently measured by ATLAS for 2.76 A TeV Pb+Pb collisions~\cite{Jia:2012sa}. It was found that the initial-state participant-plan correlators drastically differ from the corresponding final-state event-plane correlators (in the centrality dependence or sometimes even in sign)~\cite{Qiu:2012uy}.  The nonlinear event-by-event hydrodynamic evolutions couple modes among different flow harmonics, leading to a qualitatively description of the measured event-plane correlators~\cite{Qiu:2012uy}. A quantitative reproduction of the correlation strength, together with a nice fit of other flow measurements, requires systemic tuning of initial conditions, transport coefficients, etc., which has not been done yet. However, the qualitatively descriptions of the data and the presence of dramatic character change from the initial state to final state correlations strongly support the hydrodynamic descriptions of the QGP evolution~\cite{Heinz:2013th}.

\section{The QGP viscosity at RHIC and the LHC}

\quad With the efforts from different groups, it is widely accepted that the QGP shear viscosity can be extracted from the flow data~\cite{Heinz:2013th,Gale:2013da,Teaney:2009qa,
Romatschke:2009im,Romatschke:2007mq,Song:2007fn,Dusling:2007gi,Molnar:2008xj,Bozek:2009dw,Chaudhuri:2009hj,Schenke:2010rr}~\footnote{Here we concentrate on the QGP shear viscosity. The bulk viscous effects are generally neglected during the extraction of the QGP shear viscosity. For recent progress related to the bulk viscosity, please refer to Ref.~\cite{Fries:2008ts,Torrieri:2008ip,Song:2009rh,Monnai:2009ad,Denicol:2009am,Dusling:2011fd,Noronha-Hostler:2013gga}}. Using 2+1-d viscous hydrodynamics with optical Glauber and KLN initializations, Luzum and Romatschke calculated the integrated and differential elliptic flow in 200 A GeV Au+Au collisions and compared the results with the PHOBOS and STAR data~\cite{Romatschke:2007mq}. They found that $\sim$$30\%$ uncertainties from the initial eccentricities transform into $\sim$$100\%$ uncertainties for the extracted value of the QGP shear viscosity. The hadronic evolution in~\cite{Romatschke:2007mq} was simply treated as a chemical and thermal equilibrated viscous fluid expansion with a constant $\eta/s$ as input. The bulk viscosity was neglected there.  After estimating the hadronic chemical and thermal off-equilibrium effects and further considering the bulk viscous suppression of the elliptic flow, one concludes that the specific QGP shear viscosity $(\eta/s)_{QGP}$ can not exceed an upper limit at $5\times\frac{1}{4\pi}$~\cite{Romatschke:2007mq,Song:2008hj}.

\begin{figure*}[t]
\vspace*{0.0cm}
\center
\resizebox{0.90\textwidth}{4.5cm}{%
  \includegraphics{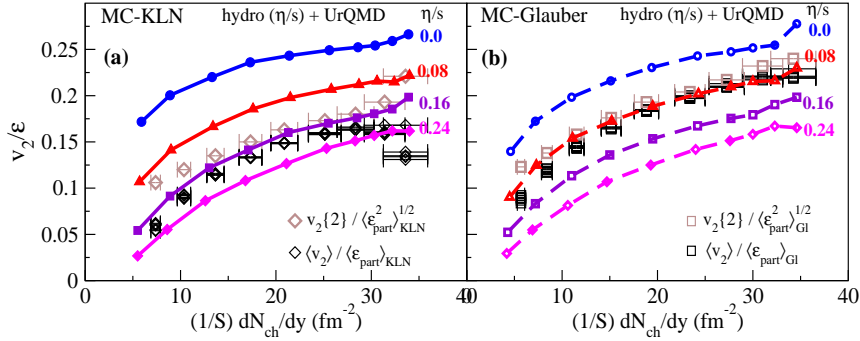}}
\vspace*{0cm}  
\caption{(Color online) eccentricity-scaled elliptic flow as a function of final multiplicity per
 area~\cite{Song:2010mg}.}
\end{figure*}

Using VISHNU hybrid model~\cite{Song:2010aq} that realistically describes the rescattering and decoupling of the hadronic matter, Ref~\cite{Song:2010mg,Song:2011hk} extracted the QGP shear viscosity from the integrated elliptic flow data in 200 A GeV Au+Au collisions. Comparing with the differential elliptic flow $v_2(p_T)$, the integrated $v_2$ for all charged hadrons is directly related to the fluid anisotropy and insensitive to the bulk viscosity, hadronic chemical compositions, the non-equilibrium particle distribution $\delta f$, etc.,  providing a robust constrain of the QGP shear viscosity. Fig.~1 shows a comparison of the eccentricity-scaled integrated elliptic flow obtained from theory and experiment. The VISHNU results are calculated with event averaged smooth initial conditions generated from MC-Glauber and MC-KLN models. Correspondingly, the experimental data used there are the corrected $v_2$ with non-flow and fluctuation effects removed~\cite{Ollitrault:2009ie}. From Fig. 1, one finds
$\frac{1}{4\pi}<(\eta/s)_{QGP}< 2.5\times\frac{1}{4\pi}$, where the main uncertainties come from these two undetermined initial conditions. Comparing with the early results from Luzum and Romatschke~\cite{Romatschke:2007mq}, the accuracy of this extracted $(\eta/s)_{QGP}$ is increased by more than a factor of two due to a better control of the hadronic evolution.  The neglected ingredients include: the effects from bulk viscosity, the event-by-event simulations, the pre-equilibrium dynamics and so on.  Due to the critical slowing down near the phase transition, the bulk viscosity only slightly suppresses the integrated $v_2$, which brings less than 20\% contaminations for the extracted QGP shear viscosity~\cite{Song:2009rh}. A comparison between event-by-event and single-shot hydrodynamics showed that for the same initial eccentricity, the $v_2$ reduction from event-by-event simulations is about 5\%. Meanwhile, the pre-equilibrium dynamics slightly increase $v_2$ by O(5\%). Considering the cancelations among these effects, the total uncertainty band for the extracted QGP shear viscosity may only slightly shift.  With $(\eta/s)_{QGP} \simeq (1/4\pi)$ for MC-Glauber and $(\eta/s)_{QGP} \simeq (2/4\pi)$ for MC-KLN read from Fig.~1, VISHNU yields an excellent description of the $p_T$-spectra and differential elliptic flow $v_2(p_T)$  for all charged and identified hadrons at various centralities in 200 A GeV Au+Au collisions~\cite{Song:2011hk}.

After extrapolating the VISHNU calculation to 2.76 A TeV Pb+Pb collisions, the $(\eta/s)_{QGP}$ extracted from RHIC slightly over-predicts the elliptic flow at the LHC~\cite{Song:2011qa}. With $(\eta/s)_{QGP}$ increased to $\sim 2.5/(4\pi)$ for MC-KLN, a better description of the integrated and differential elliptic flow for all charged hadrons has been achieved~\cite{Song:2011qa}. Comparing with the pure hydrodynamic calculations~\cite{Shen:2011eg},  VISHNU improves the description of the proton $v_2$ from most central to semi-central collisions and nicely describes the $v_2(p_T)$ mass-ordering among pions, kaons and protons at various centralities~\cite{Song:2013qma,Heinz:2011kt}. With the same inputs, it also roughly describes the $v_2(p_T)$ for $\Lambda$, $\Xi$ and $\Omega$ measured by the ALICE collaboration~\cite{Song:2013tpa}.

The initial eccentricity $\varepsilon_2$ from MC-KLN is about 30\% larger than the one from MC-Glauber. After the hydrodynamic evolution, it leads to $\sim$30\% larger elliptic flow.  As a results, the extracted specific QGP shear viscosity from MC-KLN is about twice larger than the one from MC-Glauber. Due to a similar fluctuation mechanism based on nucleon position fluctuations, the third-order eccentricities $\varepsilon_3$ from MC-KLN and MC-Glauber models are similar, leading to close triangular flow at various centralities. A fitting of the triangular flow from viscous hydrodynamics prefer a small value of the specific QGP shear viscosity $\sim$ 0.08 at the LHC~\cite{Qiu:2011hf}. With that value, MC-Glauber initial conditions nicely describe the centrality dependent elliptic and triangular flow, while MC-KLN initial conditions fail to simultaneously fit these data with an uniform QGP shear viscosity~\cite{Qiu:2011hf}. However, this does not necessarily mean the survival of the MC-Glauber model. It was found that, after the viscous hydrodynamic evolutions, both MC-Glauber and MC-KLN initial conditions fails describe the measured integrated $v_n$ (n=2...7) in ultra-central collisions by CMS and the event-by-event $v_n$ (n=2...4) distributions by ATLAS~\cite{CMS:2013bza,Jia:2012sa,Heinz,Heinz:2013wva}.

With IP-Glasma pre-equilibrium dynamics, event-by-event 3+1-d viscous hydrodynamic (MUSIC) simulations have nicely described the integrated and differential $v_n$ data with $(\eta/s)_{RHIC}=0.12$ for 200 A GeV Au+Au collisions and with $(\eta/s)_{LHC}=0.2$ for 2.76 A TeV Pb+Pb collisions~\cite{Gale:2013da,Gale:2012rq}. Impressively, the event-by-event $v_n$ distributions (n=2...4) measured by the ATLAS collaboration are also nicely described by MUSIC with $(\eta/s)_{LHC}=0.2$ for the selected centrality bins~\cite{Gale:2013da,Gale:2012rq}. Although these calculations did not couple with a hadronic afterburner, the main results will not be significantly influenced by the specific hadronic evolution since the flow at the LHC, especially for the higher order harmonics, are mainly developed in the QGP phase.

Most of the hydrodynamic simulations input a constant $\eta/s$, which corresponds to an effective specific shear viscosity averaged over the whole QGP evolution. Both the VISHNU calculations of the elliptic flow~\cite{Song:2010mg,Song:2011hk,Song:2011qa} and the event-by-event MUSIC simulations of $v_n$~\cite{Gale:2013da} showed that the averaged specific QGP shear viscosity is slightly larger at the LHC than at RHIC.  Using 3+1-d viscous hydrodynamics + UrQMD hybdrid model, a systemic fit of the elliptic flow data in 7.7, 27, 39 A GeV Au+Au collisions indicates that $(\eta/s)_{QGP} \geq 0.2$ for lower collision energies~\cite{Karpenko:2013ama}. These results demonstrate that it is possible to extract a temperature-dependent specific QGP shear viscosity $(\eta/s)_{QGP}(T)$ from the flow data measured at different collision energies. However, some issues need to be carefully investigated before a tight constrain of $(\eta/s)_{QGP}(T)$. These issues include: an EoS at finite chemical potential, the effects from net charge current and heat flow, the broken of boost-invariance and the non-equilibrium hadronic evolutions at lower collision energies, pre-equilibrium dynamics and the initialization of shear stress tenor\footnote{Ref~\cite{Shen:2010uy} showed that with an assumed $\eta/s(T)$ as input, the elliptic flow is sensitive to the initialization of the shear stress tensor at the LHC.}, etc. The extraction of $\eta/s(T)$ also requires a better determination of the initial temperature, which is closely related to the hydrodynamic starting time but can not be directly measured in experiments. Recently, Soltz and his collaborators have developed a comprehensive heavy ion model evaluation and reporting algorithm (CHIMERA) and evaluated the allowed range of initial temperature and the shear viscosity (constant $\eta/s$) from a simultaneously fitting of the pion spectra, elliptic flow, and HBT radii in 200 A GeV Au + Au collisions~\cite{Soltz:2012rk}. With the ever increasing flow data measured in experiments, it is valuable to further develop such massive data evaluating technique to systematically  evaluate different initialization models, to accurately exact the QGP shear viscosity, and to tightly constrain $\eta/s(T)$ in the future.

\section{Flow and hydrodynamics in 5.02 A TeV p+Pb collisions}

\quad In the relativistic heavy ion program at the LHC, proton-lead (p+Pb) collisions was aimed to provide reference data for Pb+Pb collisions to investigate the initial state effects. In high multiplicity p+Pb collisions at 5.02 A TeV, the measured two particle correlations between relative  pseudorapidity and azimuthal angles present ridge structures elongated in the pseudorapidity direction~\cite{CMS:2012qk,Abelev:2012ola,Aad:2013fja}. This can be explained either by the Color Glass Condensate from the initial state~\cite{Dusling:2013oia,McLerran:2013oju,Dumitru:2013tja} or by collective dynamics due to final state interactions~\cite{Bozek:2012gr}. Using two particle correlations or four-particle cumulants, the elliptic and triangular flow $v_2$ and $v_3$ have been extracted and show comparable strengthes to the ones measured in 2.76 A TeV Pb+Pb collisions~\cite{Aad:2013fja,Chatrchyan:2013nka,ABELEV:2013wsa}. Recently, the ALICE collaborations measured the $p_T$ dependent $v_2$ for identified hadrons in high multiplicity p+Pb collisions and observed a mass-splitting among pions, kaons and protons qualitatively similar to that measured in 2.76 A TeV Pb+Pb collisions~\cite{ABELEV:2013wsa}. This shows another character of the collective expansion. The $v_2(p_T)$ splitting between light and heavy particles is sensitive to the EoS and reflects the interplay of radial and anisotropy flow during the late hadronic evolution~\cite{Huovinen:2001cy}.

All of these flow data, together with the multiplicities, mean $p_T$ of identified hadrons, etc.~\cite{Abelev:2013haa}, have been semi-quantitatively described or predicted by 3+1-d hydrodynamic simulations from several groups~\cite{Bozek:2012gr,Bozek:2011if,Bozek:2013ska,Bzdak:2013zma,Qin:2013bha,Werner:2013ipa}. The triangular flow data disfavor the pure initial state descriptions of the Color Glass Condensate, which also can not directly describe/predicted the identified hadron data without additional assumptions or being combined with other models~\cite{Dusling:2013oia,McLerran:2013oju,Dumitru:2013tja}. These facts strongly indicate that, due to final state interactions, large collective flow has been generated in a smaller p+Pb system created in several TeV collisions. Instead of ruling-out the initial state scenario, it is worthwhile to evaluate the development of collective flow in different stages using a super hybrid model that combines pre-equilibrium dynamics (e.g., IP-Plasma) with hydrodynamics and hadron cascade.  Before accepting the the hydrodynamic description of a small system with large gradients, it is important to quantify the viscous effects to evaluate the applicabilities of viscous hydrodynamics. For an integrated understanding of the collective phenomena in small systems, the flow in peripheral Au+Au and Pb+Pb collisions at RHIC and the LHC should be precisely measured and re-evaluated with efforts from both experimental and theoretical sides.

\section{Summary and outlook}
\quad Viscous hydrodynamics and its hybrid models are successful tools to describe and predict the soft particle data in relativistic heavy ion collisions at RHIC and the LHC. Systematic studies of the flow data in 200 A GeV Au+Au collisions and in 2.76 A TeV Pb+Pb collisions have established that the specific QGP shear viscosity $(\eta/s)_{QGP}$ is small, not more than several times of $\frac{1}{4\pi}$. The effective (averaged) $(\eta/s)_{QGP}$ is slightly larger at the LHC than at RHIC. For 5.02 A TeV p+Pb collisions, hydrodynamics have semi-quantitatively described the flow data, which strongly indicates the generation of large collective flow. Before a quantitatively extraction of the QGP shear viscosity in 5.02 A TeV p+Pb collisions, it is necessary to further evaluate the applicabilities of hydrodynamics for the small systems created at TeV energies.

The QGP created at RHIC and the LHC are strongly coupled, its hydrodynamic evolution transform the initial state fluctuations into final state fluctuations and correlations that have been extensively measured in experiments. Some of the observables, such as the event-by-event $v_n$ distributions, indicate the importance of color charge fluctuations and may exclude these models with only nucleon fluctuations. However, a tight constrain of the initialization models, together with an accurate extraction of $(\eta/s)_{QGP}$,  requires systematic evaluations of the flow data at different aspects. These flow data include: integrated and differential harmonic flow $v_n$, flow power spectrum $v_n$ in ultra-central collisions, the event-by-event $v_n$ distribution, the event plan correlations among flow angles, the $p_T$ spectra and flow for identified hadrons, etc. To meet the above goal, event-by-event simulations of the hybrid hybrid model (that combines pre-equilibrium dynamics, hydrodynamic evolution and hadronic afterburner) and the massive data evaluating algorithm should be further developed, many of the flow data should be measured with increased accuracy.  With these related progresses becoming available in the future, it is even possible to extract the temperature-dependent QGP shear viscosity $(\eta/s)_{QGP}(T)$, bulk viscosity and other transport coefficients from the precise flow data at various collision energies. \\

{\it Acknowledgments:} This work was supported by the new faculty startup funding from Peking University.

\end{document}